\documentclass[a4paper,11pt]{article}
\usepackage{pos}

\title{Studying High-Mass Microquasars with HAWC}
 \ShortTitle{HMMQ Study with HAWC}

\author*[a]{Chang Dong Rho}
\author[b]{Ke Fang}
\author[c]{Se Yeon Hwang}
\author[c]{Youngwan Son}

\affiliation[a]{University of Seoul, Natural Science Research Institute\\
Seoul, Republic of Korea}

\affiliation[b]{University of Wisconsin-Madison, Department of Physics\\
Madison, WI, USA}

\affiliation[c]{University of Seoul, Department of Physics\\
Seoul, Republic of Korea}

\forColl{HAWC} 

\emailAdd{cdr397@uos.ac.kr}

\abstract{High-mass microquasars (HMMQs) are powerful particle accelerators, but their mechanism of the high-energy emission is poorly understood. To date, only a handful of these particle engines have ever been observed to emit gamma-ray photons and are thus potential TeV gamma-ray emitters. In this work, we study four HMMQs, namely, LS 5039, Cyg X-1, Cyg X-3, and SS 433 using the data from the High Altitude Water Cherenkov (HAWC) observatory. We perform time dependent analyses on each HMMQ to look for any periodic variations in their flux. We produce light curves using the HAWC daily maps from which we generate Lomb-Scargle periodograms. By analysing the significance of the periodogram peaks, we assess whether or not HAWC is sensitive to orbitally modulating TeV gamma-ray flux in the four HMMQs.}

\FullConference{37$^{\rm{th}}$ International Cosmic Ray Conference (ICRC 2021)\\
		July 12th -- 23rd, 2021\\
		Online -- Berlin, Germany}

\begin{document}
\maketitle

\section{Introduction}

Gamma-ray binaries are sources that have non-thermal emission peaking above 1~MeV in their spectra with orbital modulation in their gamma-ray flux \cite{dubus}. Consequently, very-high-energy (VHE) gamma-ray binaries are those that exhibit periodic variation in their VHE gamma-ray flux and are extremely rare to find. Due to the limited number of observations and lack of a standard model, our understanding of the mechanisms that drive VHE gamma-ray binaries is still poor. One of the models that exists is high-mass microquasar (HMMQ) where the ``high-mass'' refers to the massive companion star. Microquasars are a type of compact binaries that have either a neutron star or black hole as their compact object and they have materials from their massive companion that accrete onto the compact object forming an accretion disk. Microquasars also have relativistic jets ejected to either side of the disk, a feature that is also seen in active galactic nuclei or quasars, only on smaller scales.
The reasons for VHE gamma-ray flux modulation is also not clearly understood because the VHE gamma-ray binaries do not always follow the same pattern \cite{dubus}. However, to briefly discuss some of the factors, anisotropy in Compton effect would cause higher flux when the photon electron collision is head on with respect to the observer. Secondly, assuming the accretion rate and the jets contribute to the production of gamma-ray emission, the wind accretion rate is inversely proportional to the relative velocity of the compact object, which could vary with orbital position. Lastly, maximum flux from VHE gamma-ray pair production absorption would occur when the gamma-photon interaction is head-on \cite{dubus}.

The results provided in this work follow up on ``HAWC Search for High-mass Microquasars'' \cite{hmmq}, which presents TeV gamma-ray flux upper limits on the HMMQs within the HAWC field of view and discusses two scenarios that may explain potential VHE gamma-ray production. This proceeding extends on \cite{hmmq} by looking at the time-dependent aspect of the same HMMQs.

\section{HAWC}
\label{sec:hawc}

The High Altitude Water Cherenkov (HAWC) Observatory is a ground based particle sampling array designed to indirectly observe gamma rays. In order to efficiently collect particles produced from gamma-ray air showers, HAWC is located at a high altitude of approximately 4.1~km between Sierra Negra and Pico de Orizaba in Puebla, Mexico. The main array of HAWC consists of 300 water Cherenkov detectors (WCDs) covering a combined geometrical area of $\sim22,000$~m$^{2}$. Each WCD is 4.5~m tall with 7.3~m diameter and is filled with $\sim200,000$~L of purified water. At the bottom of each WCD, there are four photomultiplier tubes (PMTs) anchored so when charged daughter particles produced from air showers propagate through the WCD, the emitted Cherenkov light can be collected for indirect detection of gamma rays. Also, the WCDs are sufficiently deep enough to act as calorimeters as well as the detector medium. For each triggered event, HAWC uses the PMT charge spatial distribution to identify the ``core'' of the shower and uses the timing information to reconstruct the incidence angle of the original gamma-ray photon and thus the position of its source. The typical trigger rate of HAWC is approximately 25~kHz. However, most of the triggered events are cosmic ray background that are later filtered out using gamma-hadron separators \cite{crab}. The official energy range of HAWC is between 300~GeV and 100~TeV.

Some of the main advantages of HAWC are its high duty cycle ($>95\%$) and wide field of view (2~sr) \cite{crab}. These aspects allow HAWC to be suitable for carrying out time dependent analyses on binary systems.

The data used for the analysis presented in this proceeding are HAWC daily maps, each with one-transit (one day) worth of data, spanning from 2014 November 26th to 2020 October 7th.


\section{Results}

\begin{figure}[!htb]
  \centering
  \includegraphics[width=1.0\textwidth]{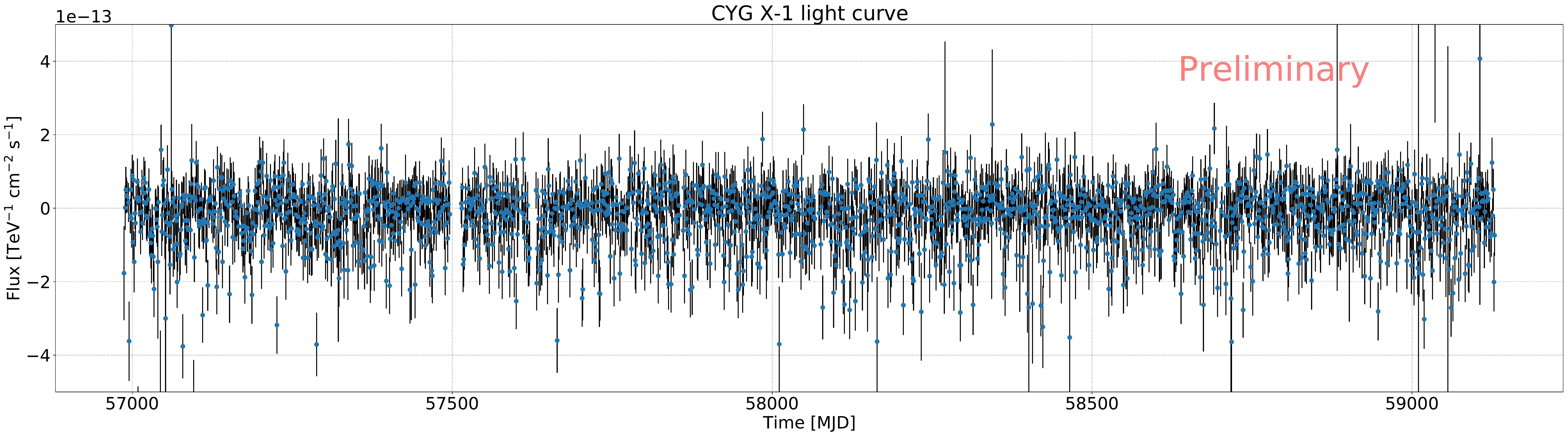}
  \includegraphics[width=1.0\textwidth]{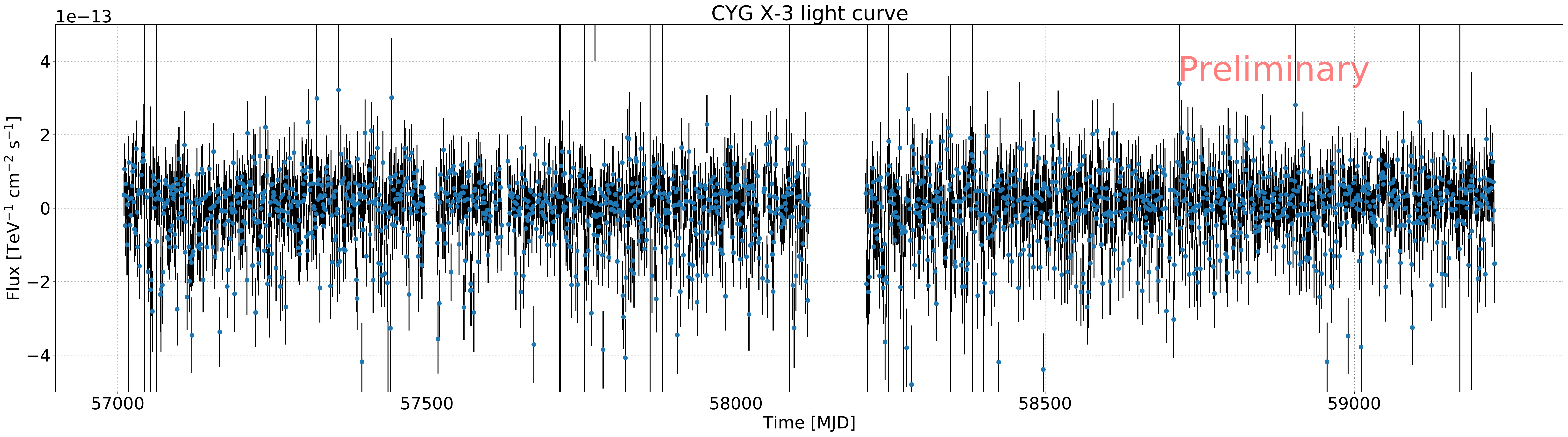}
  \includegraphics[width=1.0\textwidth]{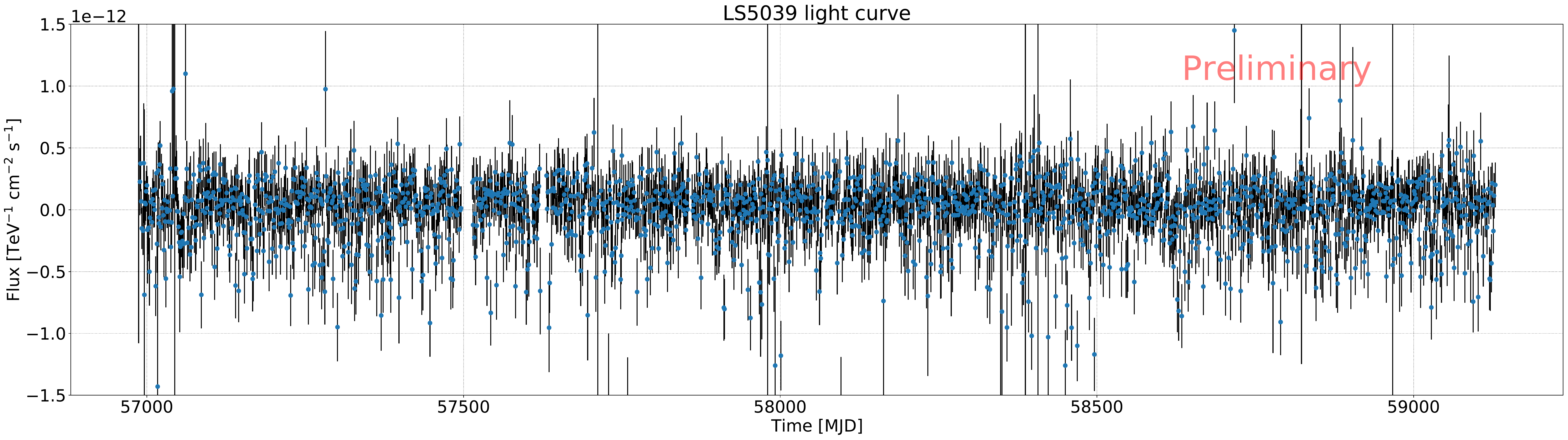}
  \includegraphics[width=1.0\textwidth]{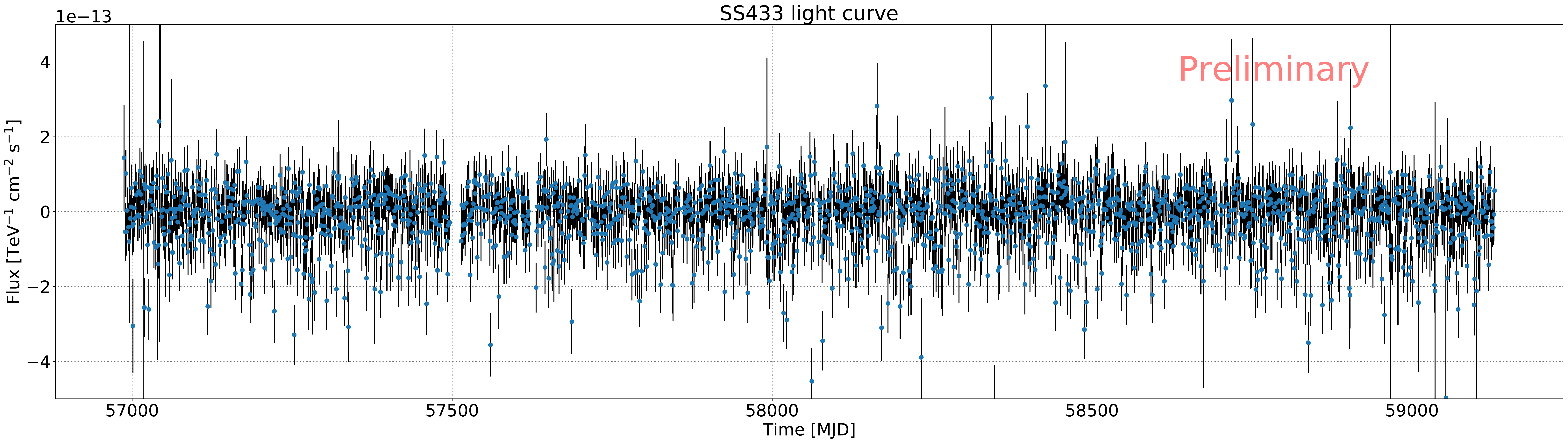}
  \caption{\sl Daily light curves for the four HMMQs, from top to bottom: Cyg X-1; Cyg X-3; LS 5039; and SS 433, produced using HAWC data.
  \label{fig:lightcurves}}
\end{figure}

Here, we present the results. The first set are the daily light curves for the four HMMQs as shown in Figure~\ref{fig:lightcurves}. Each data point corresponds to the fit result on each HAWC daily map as described in Section~\ref{sec:hawc}.

\begin{figure}[!htb]
  \centering
  \includegraphics[width=1.0\textwidth]{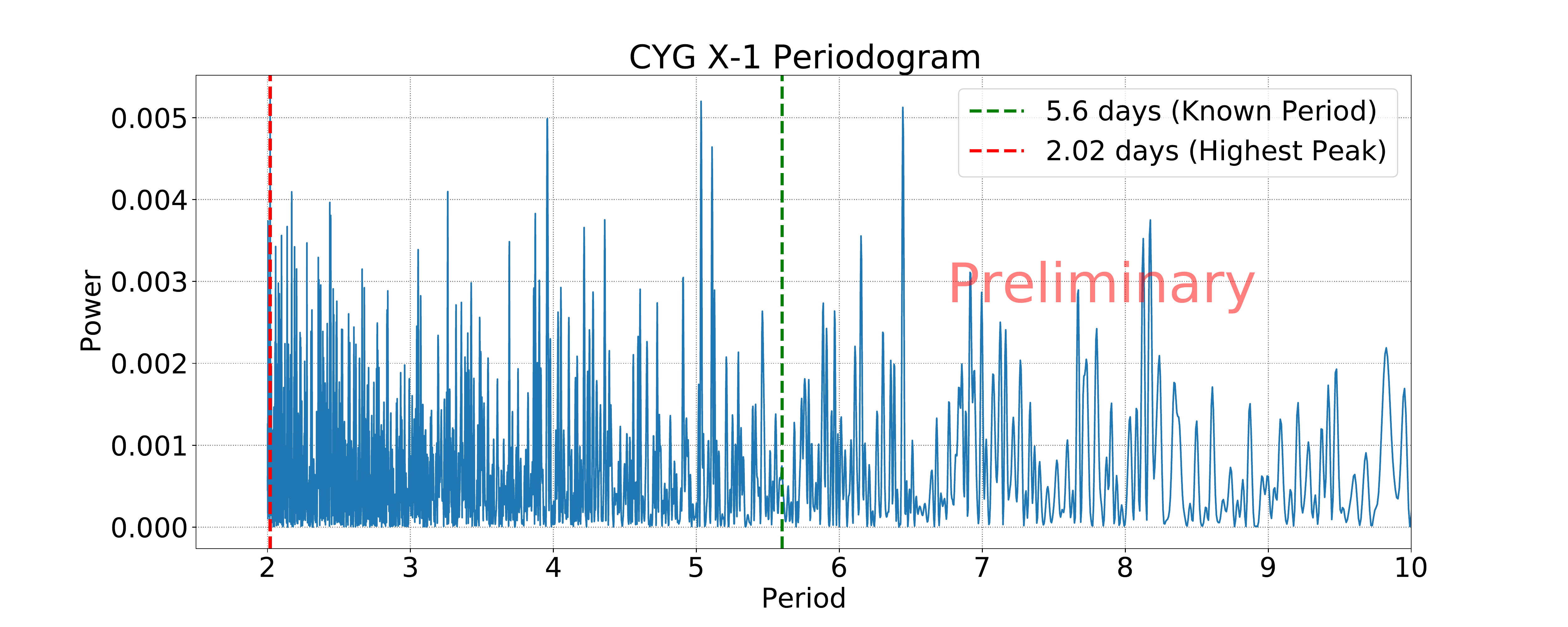}
  \includegraphics[width=1.0\textwidth]{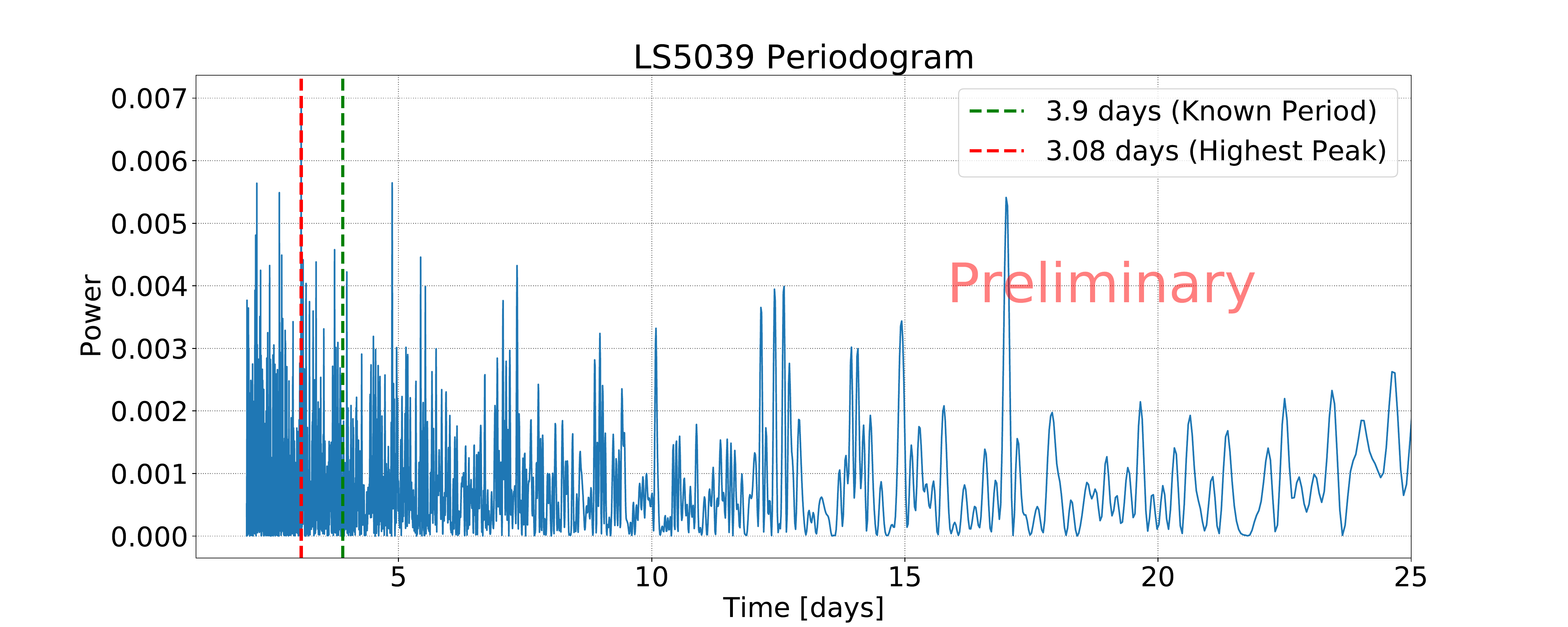}
  \includegraphics[width=1.0\textwidth]{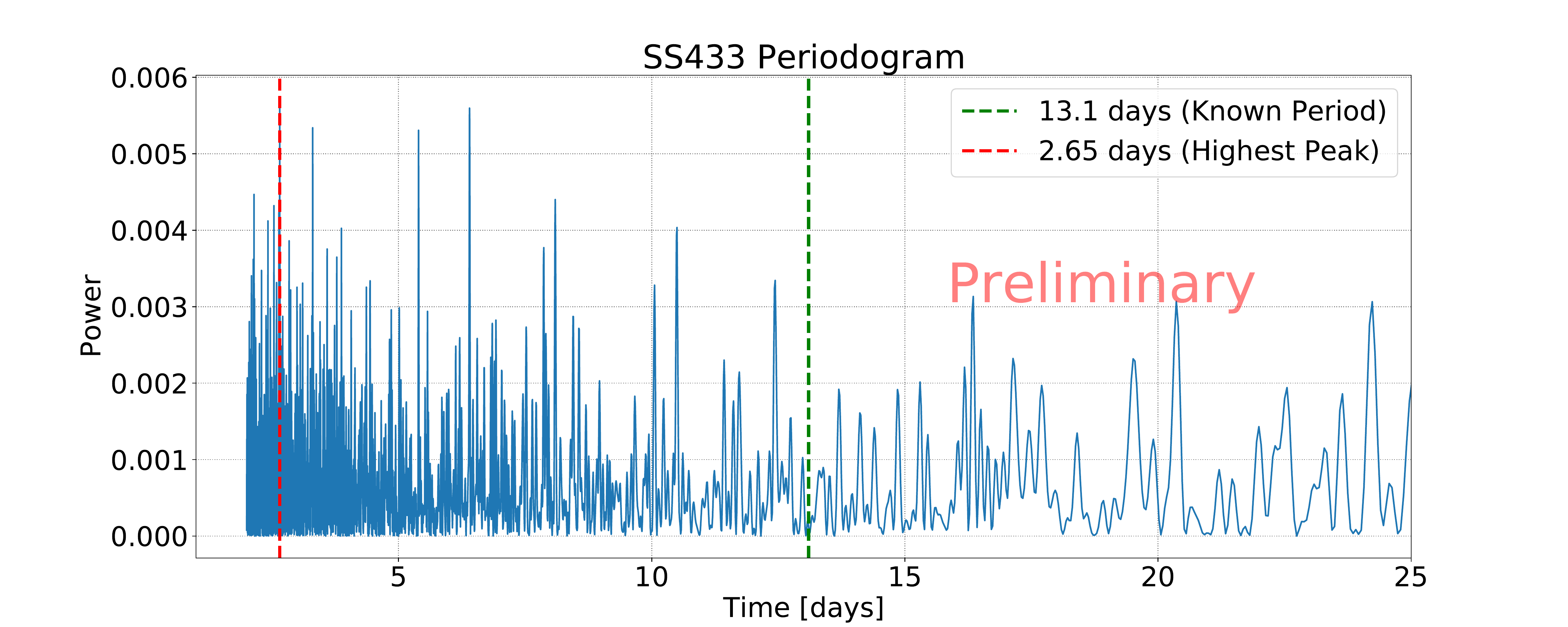}
  \caption{\sl Lomb-Scargle periodograms generated from the light curves in Figure~\ref{fig:lightcurves}. From top to bottom, these are for Cyg X-1, LS 5039, and SS 433. The green dashed lines indicate the known orbital periods of the HMMQs. The red dashed lines are mark where the maximum periodogram peaks are found. 
  \label{fig:periodograms}}
\end{figure}

Since the four HMMQs are all located in source confused regions and that the daily maps each have very small portions of the full HAWC data, studying light curves alone may not easily identify periodic orbital modulation in flux. Therefore, we applied the Lomb-Scargle periodogram \cite{lomb, scargle} to further analyse the light curves and their potential periodicities. Figure~\ref{fig:periodograms} shows the periodograms corresponding to the light curves (Figure~\ref{fig:lightcurves}). Here, the green dashed lines indicate each of their known periods: 5.6 days for Cyg X-1 \cite{cygx1}; 3.9 days for LS 5039 \cite{ls5039}; and 13.1 days for SS 433 \cite{ss433}. The red dashed lines, on the other hand, indicate the highest peaks found from the periodograms.

The periodogram for Cyg X-3 is missing because the known period for Cyg X-3 is 0.2 days \cite{cygx3}, which is shorter than the minimum time resolution of the HAWC data used (1 day), thus the periodogram analysis cannot be performed properly for Cyg X-3.

\section{Discussion}

Figure~\ref{fig:periodograms} shows that there is no significant peak at or close to the known periods of the HMMQs we are studying. As for the highest peaks, the significance of the peaks need to be computed to see if HAWC is actually observing the flux variations in HMMQs at these other periods because these peaks could simply be due to background fluctuations. To confirm, we created 10,000 pseudo light curves with background-only data for each HMMQ and made a periodogram for each light curve. Then, the power corresponding to the highest peak of each periodogram is collected to form a histogram. By looking at the power distribution on the histogram, we can compute the p-value for the original highest peak. The histograms are presented in Figure~\ref{fig:periodograms}, which show that the three HMMQs studied had the maximum peak that had a significantly small p value.

\begin{figure}[!htb]
  \centering
  \includegraphics[width=0.85\textwidth]{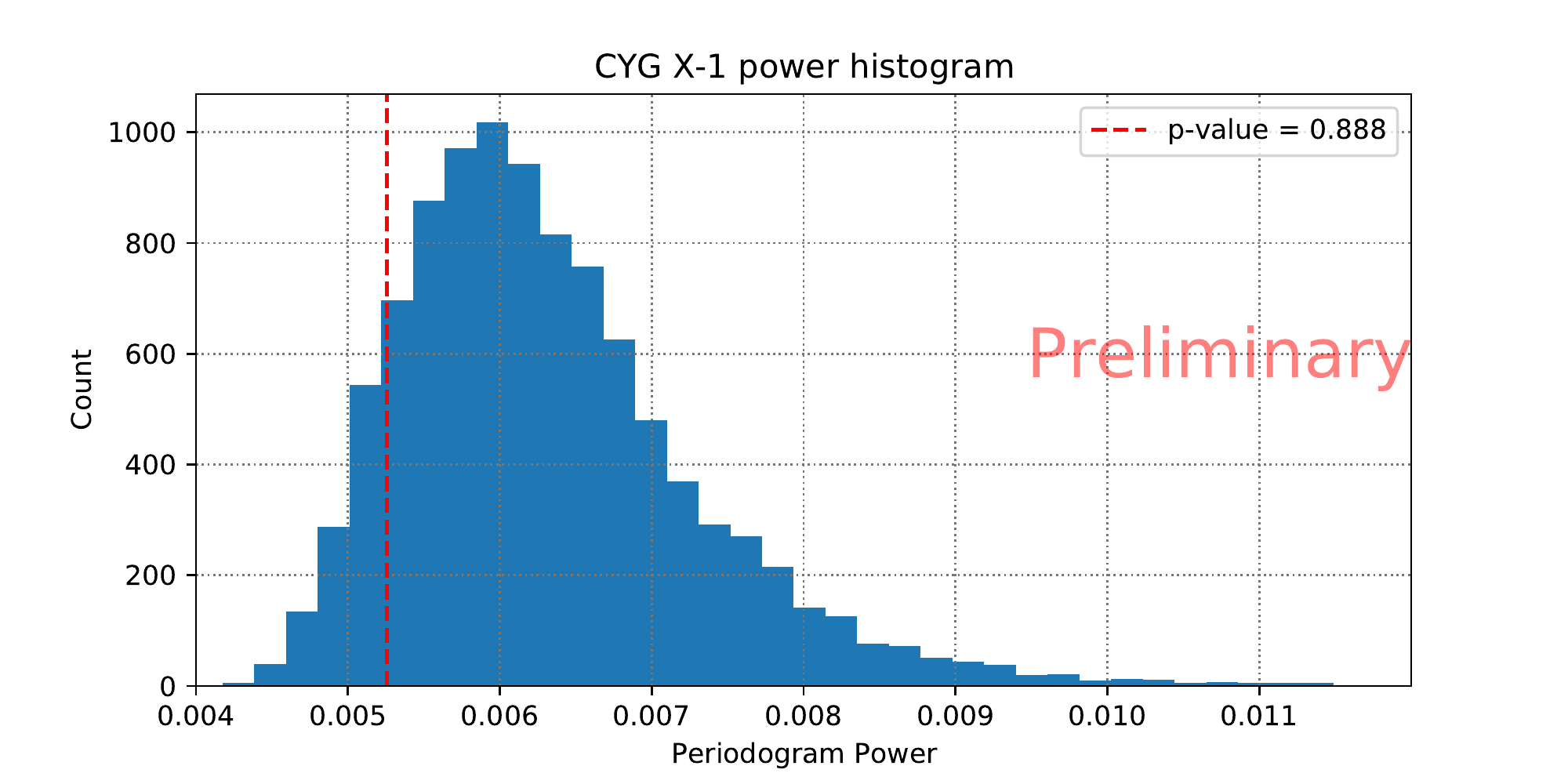}
  \includegraphics[width=0.85\textwidth]{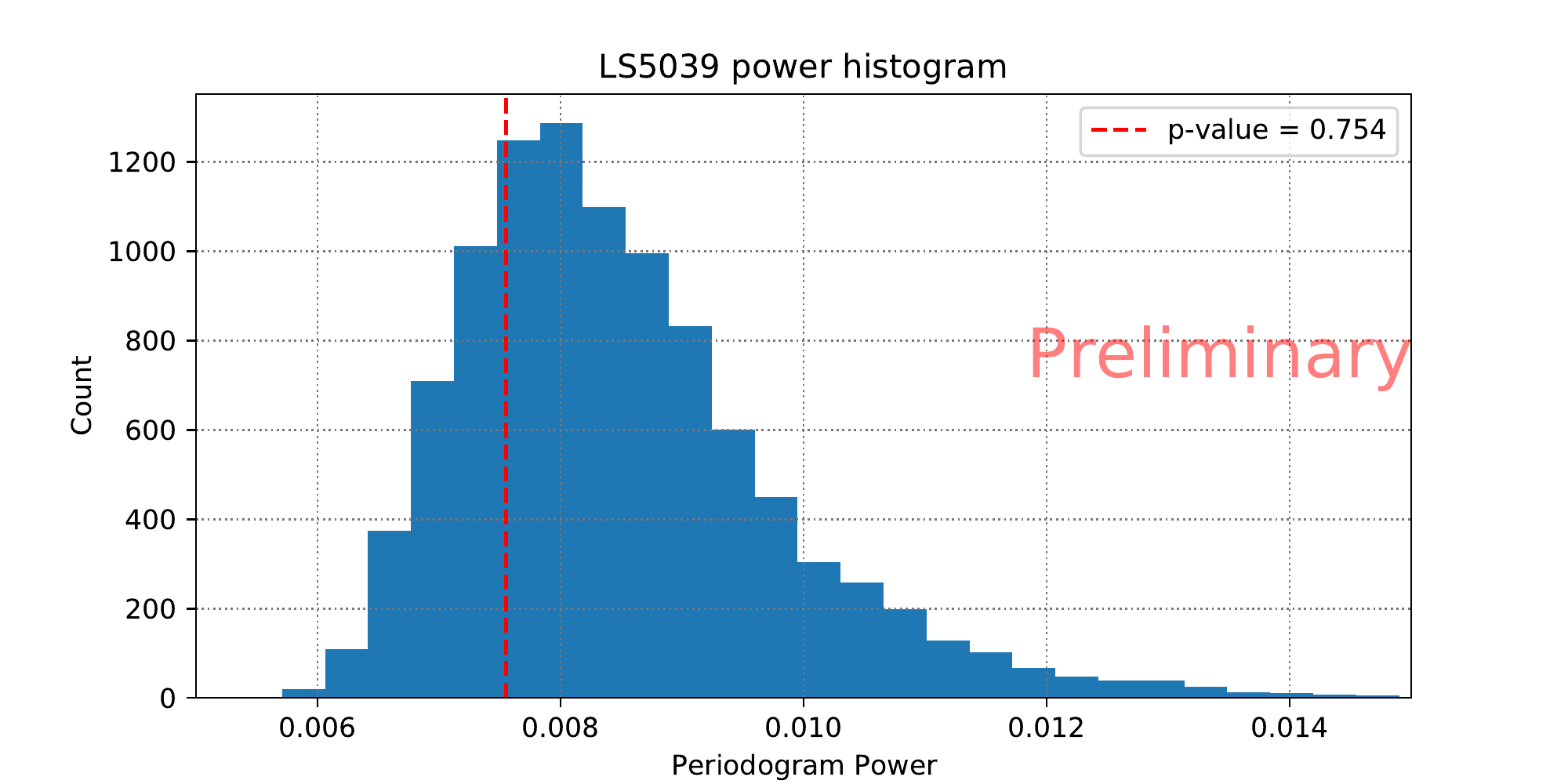}
  \includegraphics[width=0.85\textwidth]{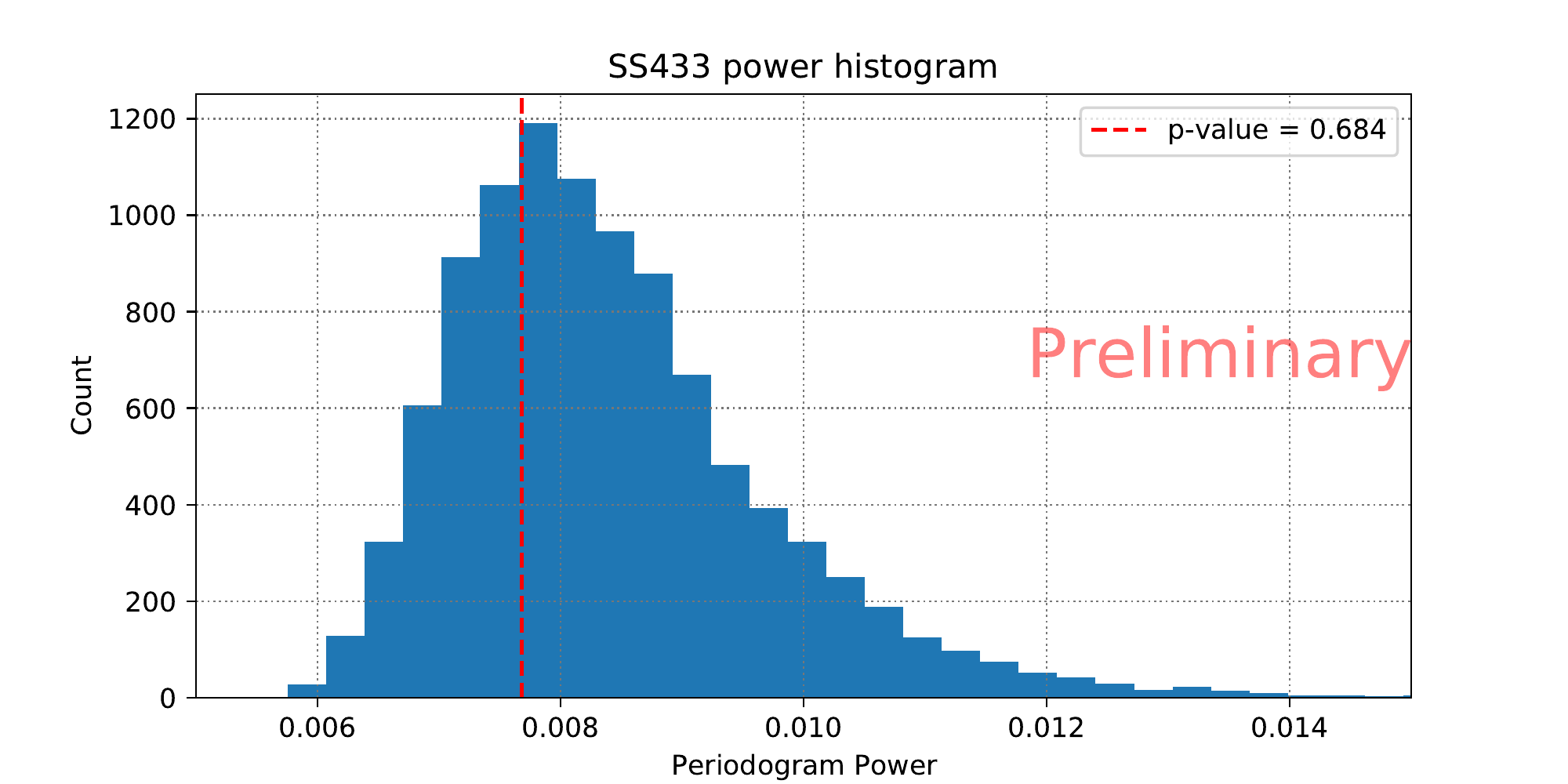}
  \caption{\sl Peak power histograms produced from 10,000 pseudo-light-curves and their periodograms. The red dashed lines indicate the p-values corresponding to the peak power from the original periodograms (Figure~\ref{fig:periodograms}).
  \label{fig:cygx1_h}}
\end{figure}

As mentioned before, LS 5039 is the one HMMQ that has been detected via TeV gamma rays \cite{hessls5039}. Through this detection, H.E.S.S. showed that this source has high and low states. Hard x-ray data also show that the HMMQs have high and low states. Assuming that there is correlation between the hard x-ray emission and VHE gamma-ray emission, collecting the parts of the HAWC data corresponding only to the high or hard states may provide the required data for HAWC to be sensitive enough to observe periodic emission. Hence, we plan to continue the analysis by concentrating on different parts of the data, more useful for each source. The Lomb-Scargle periodogram can analyse data with nonuniform time spacing, hence both time-dependent and time-independent analysis can be performed on the high-state only data. On the other hand, the high states of LS 5039 show a hard cut off above a few TeV \cite{hessls5039}. Therefore, depending on the cut-off energy, HAWC may not be able to observe the HMMQs even with the high-state data.

Another problem is that the four HMMQs are located in source confused regions with Galactic diffuse emission contamination. Therefore, multi-source fitting is preferred.

\section{Conclusion}

We have carried out a follow-up time dependent study on the four HMMQs, namely, LS 5039, Cyg X-1, Cyg X-3, and SS 433. We looked at their light curves and periodograms to look for any signs of periodically varying flux due to an orbital motion of each binary system. However, using one-day transit HAWC maps, no significant evidence could be found. We plan to continue studying the HMMQs by looking at only the part of the data that correspond to X-ray hard states and doing multi-source fits of the regions of interest. 

\acknowledgments{We acknowledge the support from: the US National Science Foundation (NSF); the US Department of Energy Office of High-Energy Physics; the Laboratory Directed Research and Development (LDRD) program of Los Alamos National Laboratory; Consejo Nacional de Ciencia y Tecnolog\'ia (CONACyT), M\'exico, grants 271051, 232656, 260378, 179588, 254964, 258865, 243290, 132197, A1-S-46288, A1-S-22784, c\'atedras 873, 1563, 341, 323, Red HAWC, M\'exico; DGAPA-UNAM grants IG101320, IN111716-3, IN111419, IA102019, IN110621, IN110521; VIEP-BUAP; PIFI 2012, 2013, PROFOCIE 2014, 2015; the University of Wisconsin Alumni Research Foundation; the Institute of Geophysics, Planetary Physics, and Signatures at Los Alamos National Laboratory; Polish Science Centre grant, DEC-2017/27/B/ST9/02272; Coordinaci\'on de la Investigaci\'on Cient\'ifica de la Universidad Michoacana; Royal Society - Newton Advanced Fellowship 180385; Generalitat Valenciana, grant CIDEGENT/2018/034; Chulalongkorn University’s CUniverse (CUAASC) grant; Coordinaci\'on General Acad\'emica e Innovaci\'on (CGAI-UdeG), PRODEP-SEP UDG-CA-499; Institute of Cosmic Ray Research (ICRR), University of Tokyo, H.F. acknowledges support by NASA under award number 80GSFC21M0002. We also acknowledge the significant contributions over many years of Stefan Westerhoff, Gaurang Yodh and Arnulfo Zepeda Dominguez, all deceased members of the HAWC collaboration. Thanks to Scott Delay, Luciano D\'iaz and Eduardo Murrieta for technical support.}

\clearpage
\section*{Full Authors List: \Coll\ Collaboration}


\scriptsize
\noindent
A.U. Abeysekara$^{48}$,
A. Albert$^{21}$,
R. Alfaro$^{14}$,
C. Alvarez$^{41}$,
J.D. Álvarez$^{40}$,
J.R. Angeles Camacho$^{14}$,
J.C. Arteaga-Velázquez$^{40}$,
K. P. Arunbabu$^{17}$,
D. Avila Rojas$^{14}$,
H.A. Ayala Solares$^{28}$
% ...

\noindent
% ...
$^{14}$Instituto de Física, Universidad Nacional Autónoma de México, Ciudad de México, México,
% ...
$^{17}$Instituto de Geofísica, Universidad Nacional Autónoma de México, Ciudad de México, México,
% ...
$^{21}$Physics Division, Los Alamos National Laboratory, Los Alamos, NM, USA,
% ...
$^{28}$Department of Physics, Pennsylvania State University, University Park, PA, USA,
% ...
$^{40}$Instituto de Física y Matemáticas, Universidad Michoacana de San Nicolás de Hidalgo, Morelia, Michoacán, México,
$^{41}$FCFM-MCTP, Universidad Autónoma de Chiapas, Tuxtla Gutiérrez, Chiapas, México,
% ...
$^{48}$Department of Physics and Astronomy, University of Utah, Salt Lake City, UT, USA,
% ...

\end{document}